\documentclass[aps,prl,superscriptaddress,twocolumn]{revtex4-1}
\usepackage{graphicx}
\usepackage{amsmath}
\usepackage{amssymb}
\usepackage{amsfonts}
\usepackage{dcolumn}
\usepackage{epsfig}
\usepackage{subfigure}
\usepackage{bm}

\begin{document}
\title{A hybrid soliton-based system: generation and steering of cavity solitons by means of photorefractive soliton electro-activation}

\author{Lorenzo Columbo}
\email{lorenzo.columbo@gmail.com} \address{Dipartimento di Scienza e Alta Tecnologia, Universit\`a  dell'Insubria, Via Valleggio 11, I-22100
Como, Italy} \address{Consiglio Nazionale delle Ricerche, CNR-IFN, I-70126 Bari, Italy}
\author{Carlo Rizza}
\address{Dipartimento di Scienza e Alta Tecnologia, Universit\`a dell'Insubria, Via Valleggio 11, I-22100 Como, Italy}
\address{Consiglio Nazionale delle Ricerche, CNR-SPIN, I-67100 Coppito L'Aquila, Italy}
\author{Massimo Brambilla}
\address{Consiglio Nazionale delle Ricerche, CNR-IFN, I-70126 Bari, Italy}
\address{Dipartimento Interateneo di Fisica, Politecnico di Bari, via Amendola 173, I-70126 Bari, Italy}
\author{Franco Prati}
\address{Dipartimento di Scienza e Alta Tecnologia, Universit\`a dell'Insubria, Via Valleggio 11, I-22100 Como, Italy}
\address{CNISM, Research Unit of Como, Via Valleggio 11, I-22100 Como, Italy}
\author{Giovanna Tissoni}
\address{Institut Non Lin\'eaire de Nice, CNRS, Universit\`e de Nice Sophia Antipolis, UMR 7335, 1361 Route des Lucioles, F-06560 Valbonne,
France}

\date{\today}
\begin{abstract}
We propose a hybrid soliton-based system consisting of a centrosymmetric photorefractive crystal, supporting photorefractive solitons, coupled
to a vertical cavity surface emitting laser, supporting multistable cavity solitons. The composite nature of the system, which couples a
propagative/conservative field dynamics to a stationary/dissipative one, allows to observe a more general and unified system phenomenology and
to conceive novel photonic functionalities. The potential of the proposed hybrid system becomes clear when investigating the generation and
control of cavity solitons by propagating a  plane wave through electro-activated solitonic waveguides in the crystal. By changing the
electro-activation voltage of the crystal, we prove that cavity solitons can be turned on and shifted with controlled velocity across the device
section. The scheme can be exploited for applications to optical information encoding and processing.
\end{abstract}
\maketitle

The study and application of self-confined structures in the coherent field of nonlinear optical systems (and in the media supporting them) is
an extremely broad and multidisciplinary field, wherein a widespread separation has consolidated between conservative and dissipative solitons.

Conservative solitons in the largest majority occur as non-diffracting beams propagating through a medium. They are thus intrinsically
transitory states in the hosting device, but their intrinsic properties make them quite appealing information carriers, as they have been proved
to be steerable, to interact in pairs or in clusters, with variegated modalities according to mutual phases; in fact logical gates, switches and
so on have been proposed and demonstrated across several classes of optical systems \cite{Trillo_1}.

On the other side, dissipative solitons occur as self-localized states in out-of-equilibrium, open systems and are often associated with
localization of patterned states which are stationary regimes of the optical system supporting them \cite{Akhmediev_1,Ackemann_1}. In
particular, cavity solitons (CSs) have been predicted and observed in broad-area systems where feedback, nonlinearity and dissipation/gain
concur in generating localization of patterned states coexisting with homogeneous ones \cite{Lugiato_1,Kuszelewicz_1}. They are thus
intrinsically persistent, independent, bistable structures whose field profile appears as intensity peaks which act as individually addressable
units of information. Moreover, CSs can be set in motion and steered in the cross section of the device by means of phase  or intensity
modulations encoded in the spatial profile of an external holding beam, with speed proportional to the gradient. This property is interpreted in
terms of Goldstone mode activation due to translational symmetry breaking in the system \cite{Firth_1, Maggipinto_1}. Such properties allow to
achieve manipulation of CSs, such as writing/erasing, shifting and applications to e.g. all-optical delay line and optical data processing
\cite{Pedaci_2,Pedaci_1}.

In this letter, we propose a hybrid soliton-based system coupling a conservative optical system exhibiting propagative solitons (namely a
photorefractive crystal (PRC)) \cite{Segev_1, DelRe_1} to a dissipative one where CSs are generated (namely a broad-area vertical cavity surface
emitting laser (VCSEL)) \cite{Barland_1, Brambilla_1, Spinelli_1}. The hybrid soliton-based system can provide the key to observe a more general
and unified phenomenology with fascinating and useful properties that can be not simply the sum of the properties of the constituent parts. In
order to display the potentials of the system, we show how the two elements can be interfaced at the same wavelength, with operational
intensities achievable in both stages, so that the light propagating in the PRC solitonic waveguides drives the CSs in the broad-area VCSEL. We
prove that cavity soliton addressing and switching, steering and deterministically delayed detection can be achieved with considerable
simplification with respect to usual optical control where a complex setup for pulse shaping and injection was required as well as a slow light
modulator \cite{Pedaci_2,Pedaci_1}. Here, due to the possibility of using fast electro-activation (on the scale of tens of nanoseconds)
\cite{DelRe_1, Sapiens_1} of the centrosymmetric photorefractive crystal, one can exploit at the same time the persistence, bistability and
plasticity of the semiconductor-based CSs. We believe that the fundamental step of proving the possibility and benefit of bridging the two
worlds of propagating and stationary solitons (and the more fundamental realms of conservative and dissipative optical systems) may open the
path to optical information manipulation in hybrid systems, where the bonuses of both constituting stages are compounded. To the best of our
knowledge a similar hybridization was previously proposed only in Ref.\cite{Terhalle_1}, where a broad-area VCSEL was coupled to an external
feedback cavity containing a PRC with an optically induced photonic lattice, which acts as a tunable filter to suppress unwanted spatial modes,
whereas in our scheme a modulation of the crystal's refractive index is introduced to break the translational symmetry of the system.

More precisely, the compact hybrid conservative-dissipative system we consider is realized by placing a centrosymmetric PRC supporting
photorefractive solitons in front of a broad-area VCSEL where CSs are generated, as sketched in Fig.1. A monochromatic plane wave is launched
through the PRC. By varying the voltage $V_{e}$ applied to the PRC through electrodes $e_{1}$ and $e_{2}$ a previously "imprinted"
photorefractive solitonic pattern is electro-activated and modulates the phase and intensity profile at the PRC exit. A proper choice of the
plane wave amplitude and the electro-activation potential $V_{e}$ allows for cavity soliton switch on and positioning in the VCSEL cross
section. We suppose that (as e.g. in Ref.\cite{Pedaci_1}) the motion can be monitored by a photodetector array which reads the VCSEL output.

The system we consider is described by the equations of spatio-temporal field evolution and structure localization in a centrosymmetric PRC with
bias \cite{DelRe_4,Ciatt_1}, coupled to those for a VCSEL with external driving \cite{Spinelli_1}. In the PRC the change of the space charge
density $\rho(x,z,t)$ due to optical photo-excitation and spatial redistribution is given by the charge continuity equation
\begin{equation}
\partial_t \rho = -\mu q \left[\nabla \cdot \left( N_e {\bf E}^{(SC)} \right)+\frac{K_B T}{q} \nabla^2 N_e \right],
\label{eq1}
\end{equation}
where $\nabla=(\partial_x,\partial_z)$, ${\bf E}^{(SC)}=(E_x^{(SC)},E_z^{(SC)})=-\nabla \phi$ is the space charge field and the
quasi-electrostatic potential $\phi$ satisfies the Poisson equation $\nabla^2 \phi=-\rho/(\epsilon_0 \epsilon_r)$ where $\epsilon_0$ is the
vacuum permittivity and $\epsilon_r$ is the relative static permittivity, $\mu$ and $q$ are the electron mobility and charge, $K_B$ is the
Boltzmann constant and
\begin{equation}
N_e = \frac{\beta}{2 \gamma} \left[\sqrt{\left(Q- \chi S\right)^2+4\chi Q \left(N_d/N_a\right)}-Q-\chi S \right]
\end{equation}
is the electron density, where $N_a$ and $N_d$ are the acceptor and donor impurity density, respectively. Note that $\beta$ is the rate of
thermal excitation of electrons, $\gamma$ is the electron-ionized trap recombination rate, $S=1+\rho/(q N_a)$, $\chi=\gamma N_a/\beta$,
$Q=1+|E_{PR}/E_b|^2$, $E_{PR}(x,z,t)$ is the slowly-varying amplitude of the optical electric field polarized along $x$ (at wavelength
$\lambda_{PR}$) and $E_b$ is the amplitude of the uniform background illumination. Furthermore, the optical field dynamics is described by
\begin{equation}
i \partial_z E_{PR}+\frac{\partial^2_x E_{PR}}{2k_{PR}}  =\frac{k_{PR}}{2} n_{PR}^2 g \epsilon_0^2(\epsilon_r-1)^2 [E^{(SC)}_x]^2  E_{PR},
\label{eq3}
\end{equation}
where $n_{PR}$ is the uniform background refractive index, $k_{PR}=2 \pi n_{PR}/\lambda_{PR}$ and $g$ is the effective electro-optic
coefficient. The dynamical equations describing radiation-matter interaction in the VCSEL are \cite{Spinelli_1}
\begin{eqnarray}
\partial_t E_V&=&\frac{1}{\tau_p}\Big[-\left(1+i\theta \right)E_V+E_I+2C(1-i\alpha) \nonumber \\
              &\cdot&(N-1)E_V\Big] +i \frac{c}{2 k_V n_V} \partial_x^{2}E_V , \nonumber\\
\partial_t N &=&-\frac{1}{\tau_e} \left[N-I_p+|E_V|^{2}(N-1) \right],  \label{carrierad}
\end{eqnarray}
where $E_V(x,t)$ is the slowly-varying amplitude of the optical electric field polarized along $x$ (at wavelength $\lambda_V$) and scaled to the
characteristic amplitude $E_0$, $E_I=E_{PR}(x,z=L_{PR},t)/(E_0 \sqrt{T})$, $T=1-R$ where $R$ is the mirror reflectivity, $\tau_p$ is the photon
decay time, $\theta$ is the cavity detuning between $\omega=2 \pi c/\lambda_V$ and the closest cavity resonance, $C$ is the gain-to-loss ratio,
$\alpha$ is the linewidth enhancement factor, $N$ is the carriers density scaled to the transparency value $N_{0}$, $k_V=2 \pi n_V /\lambda_V$
and $n_V$ is the background refractive index. The constant $E_0$ is associated with the saturation intensity $I_0=\epsilon_0 c n_V E_0^2/2=\hbar
\omega L_A N_0/(4 \tau_e T C)$ where $L_A$ is the length of the region filled by the active medium (generally $I_0 \sim10$ KW$/$cm$^2$). In the
carrier density equation, $I_p$ is the pump current, $\tau_e$ is the carriers density decay time and we neglect radiative decay and diffusion.
Here, we have chosen photorefractive parameters associated with a crystal sample of potassium lithium tantalate niobate (KLTN) at room
temperature \cite{Ciatt_1} and VCSEL parameters associated to a single longitudinal mode GaAs-GaAlAs laser slightly below threshold
\cite{Spinelli_1,Note_2}. Although the model in Ref.\cite{Spinelli_1} is intrinsically suited for a $2D$ device such as a VCSEL, we reduced it
to one transverse dimension, because the VCSEL is coupled with a PRC where we have previously written a $1D$ waveguide. In practice, this
amounts to consider either a VCSEL that is coherently injected along a stripe or a rectangular VCSEL \cite{Groneborn_1}.

The proposed technique for obtaining the generation and the fast control of CSs consists of a two-stage process: (i) a writing phase at the
active wavelength $\lambda_{PR}=0.5$ $\mu$m, (ii) a read-out at the non-photorefractive wavelength $\lambda_{PR}=\lambda_V=0.85$ $\mu$m and
consequently the generation and speed control of CSs by means of soliton electro-activation.

The first phase (i) consists in 'writing' the charge distribution inside the crystal; here, oppositely charged solitonic waveguides are
imprinted in the PRC. Note that, in this preliminary stage, the PRC is not coupled with the cavity so that the possible reflections from VCSEL
are absent. The 'writing' is obtained by launching in the PRC a Gaussian beam $E_{PR}(x,z=0,t)= E_b u_0 e^{-(x-x_0(t))^{2}/(2 \sigma^2)}$, where
$u_0=1.06$, $\sigma=6.1$ $\mu$m, $x_0$ is the beam position. Optimally symmetric waveguides are obtained by periodically alternating the beam
position $x_0(t)$ and the writing applied voltage $V_e(t)$ as follows: one sets $x_0= 10$ $\mu$m, $V_e= 10$ V for $0<t<T_0/2$ and $x_0=-10$
$\mu$m, $V_e=-10$ V for $T_0/2<t<T_0$ where the period $T_0=0.004 \tau$ ($\tau=\gamma \epsilon_0 \epsilon_r /(q \beta \mu)$ is the dielectric
relaxation time, generally it is of the order of minutes). The writing phase lasts $0.4 \tau$.

The second phase (ii), where the PRC is coupled to the VCSEL, is the fast modulation of the refractive index of the crystal which creates a
field landscape in the beam impinging onto the VCSEL, capable of controlling the cavity soliton. This is achieved by launching a monochromatic
plane wave with amplitude $E_{PR}(x,z=0,t)=0.77 E_0\sqrt{T}$ into the PRC. It propagates through the refractive index landscape created in the
writing phase but is does not alter it since $\lambda_V=0.85$ $\mu$m is in the near-infrared where the crystal does not show any photorefractive
effect \cite{Note_1} (as a consequence the carrier density $\rho$ does not change over time). The output field, originally uniform, then
acquires a modulation which can be controlled (as we show in the following) by varying the electro-activation voltage $V_e(t)$. In Fig.2, we
plot the refractive index variation $\delta n$ (a), the phase $\Phi_{PR}$ (b) and the normalized modulus $|E_{PR}/(E_0 \sqrt{T})|$ (c) of the
optical electric field at the PRC exit facet ($z=L_{PR}$) in the read out phase for the electro-activation potential $V_e=\pm 40$ V and $V_e=0$
V. During propagation, the originally transversely uniform near-infrared wave at $z=0$ is focussed into the guiding structure and defocussed out
of the antiguiding structure. Hence, as shown in Fig.2(b) and Fig.2(c), the solitonic waveguides electro-activation induces controllable local
maxima and minima of the field profile at the crystal exit. Different biases will tailor different landscapes at the PRC exit facet and thus
cause different, controllable effects on the VCSEL field.
\begin{figure}
\center
\includegraphics*[width=0.5\textwidth]{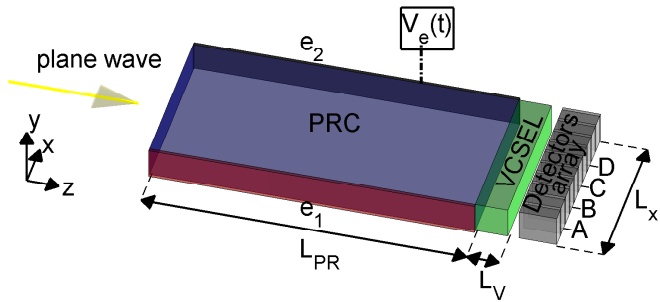}
\caption{(Color on-line) Schematic of the proposed setup. A plane wave is launched through a centrosymmetric PRC in contact with a VCSEL where
CSs are generated. The CSs location and drift velocity can be controlled by the applied voltage $V_e(t)$ (through the electrodes $e_1$ and
$e_2$) and can be monitored by a detectors array (tagged with A,B,C and D).}
\end{figure}
\begin{figure}
\center
\includegraphics*[width=0.5\textwidth]{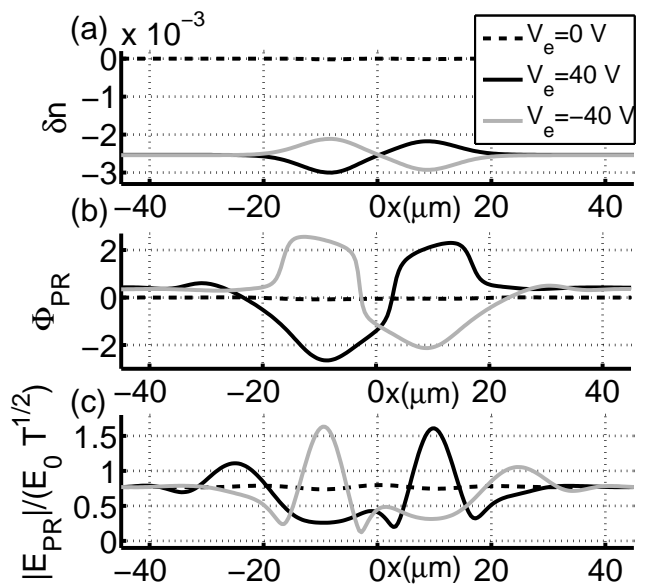}
\caption{(Color on-line) (a) Refractive index profile $\delta n$, field phase $\Phi_{PR}$ (b) and dimensionless field amplitude
$|E_{PR}/(E_0\sqrt{T})|$ (c) at $z=L_{PR}$ and during the read-out phase for $V_e=0$ V (black dashed line), $V_e=40$ V (grey solid line) and
$V_e=-40$ V (black solid line).}
\end{figure}
As reported e.g in Refs. \cite{Spinelli_1, Barland_1} there exist injected intensity ranges where a coherently driven VCSEL shows multistability
between a transversely uniform low intensity configuration and CSs. Hence, by properly adjusting the electro-activation potential $V_{e}$ and
the amplitude of the plane wave, we reach the necessary intensity modulation at the predetermined position to switch on a cavity soliton. No
addressing pulses are required, showing that electro-optical control of spatially distributed structures can be achieved by acting on global
parameters; this also allows to achieve a considerable simplification of the setup.

The time plot in Fig.3 shows the independent switch-on of two CSs. We inject into the VCSEL the field emerging from the PRC in the case where
the initial plane wave amplitude is $E_{PR}(x,z=0,t)=0.77E_0 \sqrt{T}$ with no bias ($V_{e}=0$ V; see black dashed lines in Fig. 2(b) and Fig.
2(c)). In this case the slight modulation of the holding beam phase and modulus is insufficient to generate a structure in the VCSEL. After
$\sim 70$ ns by setting $V_{e}=40$ V and consequently enhancing the holding beam modulation (see black solid line in Fig. 2(b) and Fig. 2(c)) a
cavity soliton is switched on at $x_{1a}=14$ $\mu$m. After $35$ ns we bring back to $0$ V the electro-activation potential and the cavity
soliton persists, although it moves to the new equilibrium position of $x_{2a}=32.4$ $\mu$m. After further $\sim 210$ ns, the potential is set
to $-40$ V for the same duration as before and an independent cavity soliton now appears at $x_{1b} = -7.5$ $\mu$m. The new cavity soliton
drifts to $x_{2b}=-26.3$ $\mu$m where it remains stable when the voltage is again set to $V_e^*=0$ V. The slight asymmetry with the other cavity
soliton location reflects the asymmetry of the two waveguides visible in Fig.2(b). It is worth noting that the cavity soliton location at steady
state is determined by both the phase and intensity gradients resulting at PRC exit. The effective control of CSs drift, via the tailoring of
the plane wave profile by an electro-optically activated PRC, is proved further when the final value of the voltage $V_e^*$ at the PRC is not
set to zero after cavity soliton excitation, but kept at a moderate value ranging from $V_e^* = 0.01$ V to $V_e^* = 1.05$ V. In this case a
slight intensity and phase modulation of the wave driving the VCSEL is capable of changing the final location of the cavity soliton, which
ranges from $26$ $\mu$m to $32$ $\mu$m for e.g. the right one.
\begin{figure}
\center
\includegraphics*[width=0.5\textwidth]{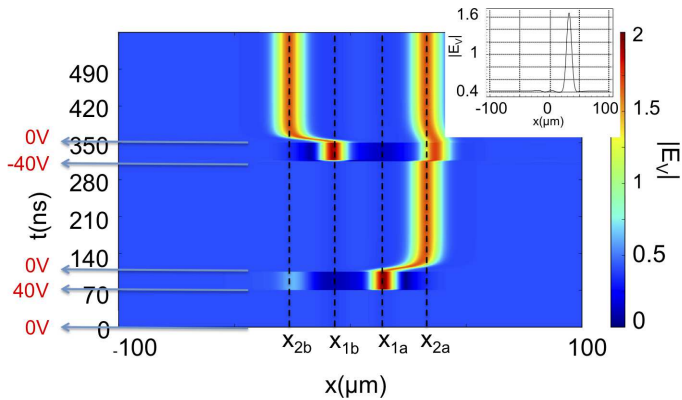}
\caption{ (Color on-line) Time plot showing the sequential and independent switch on of two stable CSs in two almost symmetric positions respect
to the device center by means of photorefractive soliton electro-activation. The arrows indicate instants where $V_{e}$ commutation occurs. In
the inset we report the one-soliton transverse profile at steady state.}
\end{figure}
\begin{figure}
\center
\includegraphics*[width=0.5\textwidth]{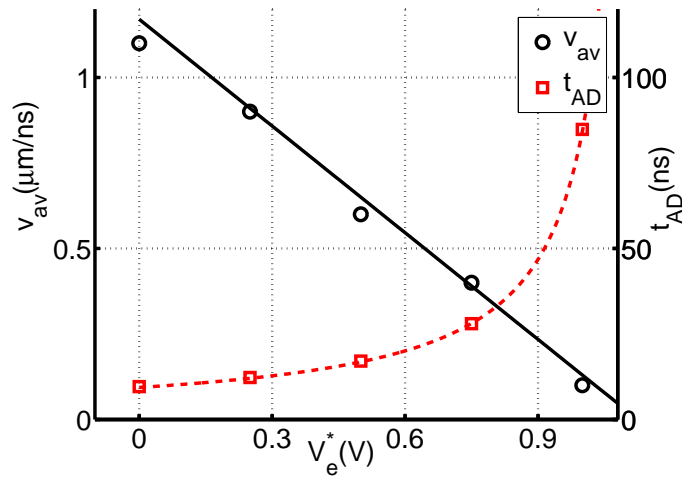}
\caption{(Color on-line) Average drift velocity $v_{av}$ (black circle) and arrival time at detector $D$ $t_{AD}$ (red square) versus $V_{e}^*$.
Continuous lines represent fitting curves.}
\end{figure}

Finally, we demonstrated that the capability of the hybrid device of modelling the refractive index profile in the PRC stage allows to control
the CSs motion and in particular its drift velocity. In our simulations, we monitored the passage times of the CSs (switched on at the location
of detector $A$ in our layout) across a fixed location (e.g. the one corresponding to detector $D$) when the final bias $V_e^*$ is kept at
different constant values. We denote as $T_{AD}$ the time in which the cavity soliton reaches a detector $D$ distant $11$ $\mu$m from $A$ (see
Fig.4). The corresponding average velocity $v_{av}=11$ $\mu$m$/t_{AD}$ is also plotted in the figure where it is evident that it depends
linearly on $V_e^*$ and it has a maximum value of $\sim 1$  $\mu$m$/$ns in excellent agreement with theoretical and experimental evidences
obtained with optical gradient modelling reported in earlier literature \cite{Pedaci_1}. Thus, once the slow writing phase has been
accomplished, the CSs can be detected with controlled delay by separate detectors. The CSs control can be performed by fast bias control and
does not require slow phase profiling or complex set-up for collimation of addressing pulses.

In conclusion, we investigate a composite soliton-based physical system which bridges propagative (conservative) photorefractive field dynamics
to its bistable (dissipative) counterpart in a microlaser. We provide a proof-of-principle demonstration of an original, high speed and
integrated device for CSs generation and manipulation in a VCSEL through fast soliton electro-activation of a centrosymmetric PRC. It is worth
noting that no addressing pulses are required, hence a considerable simplification of the setup is achieved. This scheme lends itself to all
optical delay-line or signal buffering with a high potential for applications. Finally, we believe that the considered hybrid system can show
fascinating and useful properties due to the nontrivial interaction between photorefractive solitons and CSs where either a photorefractive
soliton controls cavity soliton (or viceversa a cavity solitons control a photorefractive soliton) or the CSs and photorefractive solitons are
both capable of simultaneously influencing each other.

Authors thank Dr. Eugenio Del Re and Dr. Stephane Barland for many fruitful discussions. This research has been funded by the Italian Ministry
of Research (MIUR) through the "Futuro in Ricerca" FIRB-grant PHOCOS - RBFR08E7VA.


\begin{thebibliography}{1}
\bibitem{Trillo_1} S. Trillo and W. E. Torruellas, eds., Spatial Solitons, Springer-Verlag, Berlin (2001).
\bibitem{Akhmediev_1} N. Akhmediev, A. Ankiewicz, eds., Dissipative Solitons, Springer, Heidelberg (2005).
\bibitem{Ackemann_1} T. Ackemann, W.J. Firth, G. Oppo, Advances in Atomic, Molecular and Optical Physics {\bf 57}, 323 (2009).
\bibitem{Lugiato_1} L.A. Lugiato, IEEE J. Quantum Electron. {\bf 39}, 193 (2003).
\bibitem{Kuszelewicz_1} R. Kuszelewicz, S. Barbay, G. Tissoni and G. Almuneau, EPJD {\bf 59}, 1 (2010).
\bibitem{Firth_1} W. J.  Firth and A. J. Scroggie, Phys. Rev. Lett. {\bf 76}, 1623 (1996).
\bibitem{Maggipinto_1} T. Maggipinto, M. Brambilla, G. K. Harkness, and W. J. Firth, Phys. Rev.E {\bf 62}, 8726 (2000)
\bibitem{Pedaci_2} F. Pedaci, P. Genevet, S. Barland, M. Giudici and J. R. Tredicce, Appl. Phys. Lett. {\bf 89}, 221111 (2006).
\bibitem{Pedaci_1} F. Pedaci, S. Barland, E. Caboche, P. Genevet, M. Giudici and J. R. Tredicce, T. Ackemann, A. J. Scroggie,
                   W. J. Firth, G. L. Oppo, G. Tissoni, and R. J\"{a}ger, Appl. Phys. Lett. {\bf 92}, 011101 (2008).
\bibitem{Segev_1} M. Segev and A. J. Agranat, Opt. Lett. {\bf 22}, 1299 (1997).
\bibitem{DelRe_1} E. DelRe, M. Tamburrini, A.J. Agrant, Opt. Lett. {\bf 25}, 963 (2000).
\bibitem{Barland_1} S. Barland, J. R. Tredicce, M. Brambilla, L. A. Lugiato, S. Balle, M. Giudici,
                    T. Maggipinto, L. Spinelli, G. Tissoni, T. Kn\"{o}dl, M. Miller and R. J\"{a}ger, Nature {\bf 419}, 699 (2002).
\bibitem{Brambilla_1} M. Brambilla, L. A. Lugiato, F. Prati, L. Spinelli and W. J. Firth, Phys. Rev. Lett. {\bf 79}, 2042 (1997).
\bibitem{Spinelli_1} L. Spinelli, G. Tissoni, M. Brambilla, F. Prati and L. A. Lugiato, Phys. Rev. A {\bf 58}, 2542 (1998).
\bibitem{Sapiens_1} N. Sapiens, A. Weissbrod, A. J. Agranat, Opt. Lett. {\bf 34}, 353 (2009).
\bibitem{Terhalle_1} B. Terhalle, N. Radwell, P. Rose, C. Denz, and T. Ackemann, Appl. Phys. Lett. {\bf 93}, 151114 (2008).
\bibitem{DelRe_4} E. DelRe, A. Ciattoni, and E. Palange, Phys. Rev. E {\bf 73}, 017601 (2006).
\bibitem{Ciatt_1}  A. Ciattoni, E. DelRe, A. Marini, C. Rizza, Optics Express {\bf 16}, 10867 (2008).
\bibitem{Note_2}  In all simulations, we used the following parameters:
                  $L_x=200 $ $\mu$m, $L_{PR}=1$ mm, $L_V = 2$ $\mu$m,
                  $\chi=10^4$, $N_a=3.04 \cdot 10^{22} $ m$^{-3}$, $N_d=101 \cdot N_a$, $n_{PR}=2.4$, $g=0.13$ m$^4$C$^{-2}$ , $\epsilon_r=3 \cdot 10^4$,
                  $\alpha=5$, $\theta=-2$, $C=0.45$, $I_p=2$, $\tau_p=11.7$ ps, $n_V=3.5$, $\tau_e=1$ ns, $R=0.996$ and $L_A=50$ nm.
\bibitem{Groneborn_1} S. Gronenborn, J. Pollmann-Retsch, P. Pekarski, M. Miller, M. Str\"{o}sser,
                      J. Kolb, H. M\"{o}nch, P. Loosen, Appl. Phys B {\bf 105}, 783 (2011).
\bibitem{Note_1} Note that we have not considered time dispersion effects so that the background refractive
                 index $n_{PR}$ and the electro-optic constant $g$ does not depend on frequency $\omega$.
\end{thebibliography}
\end{document}